\documentclass[12pt,a4paper]{article}
\usepackage{a4wide}
\usepackage{amsmath}
\usepackage{amssymb}
\usepackage{epsfig}
\usepackage{subfigure}
\usepackage{exscale}
\usepackage{float}
\usepackage{bbm}
\usepackage[numbers,sort&compress]{natbib}

\newcommand{\R}{{\mathbb{R}}}
\newcommand{\C}{{\mathbb{C}}}
\newcommand{\1}{{\mathbbm{1}}}
\newcommand{\p}{\partial}

\makeatletter
\@addtoreset{equation}{section}
\makeatother

\title{Systematic Low-Energy Effective Field Theory for \\
Electron-Doped Antiferromagnets}

\author{C.\ Br\"ugger$^a$, C.\ P.\ Hofmann$^b$, F.~K\"ampfer$^a$, M.~Moser$^a$,
  M.\ Pepe$^c$,\\ and U.-J.~Wiese$^a$
\\ \\
$^a$ Institute for Theoretical Physics, Bern University \\
Sidlerstrasse 5, CH-3012 Bern, Switzerland \\ \\
$^b$ Facultad de Ciencias, Universidad de Colima \\
Bernal D\'iaz del Castillo 340, Colima C.P.\ 28045, Mexico \\ \\
$^c$ Istituto Nazionale di Fisica Nucleare and \\
Dipartimento di Fisica, Universit\`a di Milano-Bicocca \\
3 Piazza della Scienza, 20126 Milano, Italy \\ \\ }

\begin{document} 
\maketitle

\vspace{-1cm}

\begin{abstract} \normalsize

In contrast to hole-doped systems which have hole pockets centered at
$(\pm \frac{\pi}{2a},\pm \frac{\pi}{2a})$, in lightly electron-doped 
antiferromagnets the charged quasiparticles reside in momentum space pockets 
centered at $(\frac{\pi}{a},0)$ or $(0,\frac{\pi}{a})$. This has important
consequences for the corresponding low-energy effective field theory of 
magnons and electrons which is constructed in this paper. In particular, in 
contrast to the hole-doped case, the magnon-mediated forces between two 
electrons depend on the total momentum $\vec P$ of the pair. For $\vec P = 0$ 
the one-magnon exchange potential between two electrons at distance $r$ is
proportional to $1/r^4$, while in the hole case it has a $1/r^2$ dependence.
The effective theory predicts that spiral phases are absent in electron-doped 
antiferromagnets.

\end{abstract}
 
\maketitle
 
\newpage

\section{Introduction}

Although they are not yet high-temperature superconductors, understanding
lightly doped antiferromagnets is a great challenge in condensed matter 
physics. A lot is known about hole- and electron-doped systems both from 
experiments and from studies of microscopic Hubbard or $t$-$J$-type models
\cite{Bri70,Hal83,And87,Gro87,Shr88,Cha89,Kan89,Sac89,Wen89,Sha90,Mon91,Kue93,Kuc93,Fla93,Sus94,Dag94,Goo94,Bel95,Kuc95,Yam99,Bru00,Kus02,Arm02,Yam03,Sac03,Mar03,Lee03,Kus03,Sen04,Yua04,Toh04,Guo06,Aic06}.
Based on the work of Haldane \cite{Hal83} and of Chakravarty, Halperin, and 
Nelson \cite{Cha89} who described the low-energy magnon physics by a $(2+1)$-d 
$O(3)$-invariant nonlinear $\sigma$-model, several attempts have been made to
include charge carriers in the effective theory 
\cite{Shr88,Wen89,Sha90,Kue93}. However, conflicting results have
been obtained. For example, the various approaches differ in the fermion 
field content of the effective theory and in how various symmetries are 
realized on those fields. In particular, it has not yet been established that 
any of the effective theories proposed so far indeed correctly describes the 
low-energy physics of the underlying microscopic systems in a quantitative
manner. 

In analogy to chiral perturbation theory for the pseudo-Nambu-Goldstone pions 
of QCD \cite{Wei79,Gas85}, the $(2+1)$-d $O(3)$-invariant nonlinear 
$\sigma$-model has been established as a systematic and quantitatively correct 
low-energy effective field theory in the pure magnon sector
\cite{Cha89,Neu89,Fis89,Has90,Has91,Has93,Chu94,Leu94,Hof99}. 
In analogy to baryon chiral perturbation theory 
\cite{Geo84,Gas88,Jen91,Ber92,Bec99} --- the effective theory for pions and 
nucleons --- we have recently extended the pure magnon effective theory by
including charge carriers \cite{Kae05,Bru05,Bru06}. The effective theory 
provides a powerful theoretical framework in which the low-energy physics of 
magnons and charge carriers can be addressed in a systematic manner. The 
predictions of 
the effective theory are universal and apply to a large class of doped 
antiferromagnets. This is in contrast to calculations in microscopic models 
which usually suffer from uncontrolled approximations and are limited to just 
one underlying system. While some results obtained with the effective theory
can be obtained directly from microscopic systems, the effective field theory 
treatment allows us to derive such results in a systematic and more transparent
manner and it puts them on a solid theoretical basis. In order not to obscure
the basic physics of magnons and charge carriers, the effective theory has been
based on microscopic systems that share the symmetries of Hubbard or 
$t$-$J$-type 
models. In particular, effects of impurities, long-range Coulomb forces, 
anisotropies, or small couplings between different $CuO_2$ layers have so far
been neglected, but can be added whenever this becomes desirable. Before such 
effects have been included, one should be aware of the fact that the effective 
theory does not describe the actual materials in all details. Still, for 
systems that share the symmetries of the Hubbard or $t$-$J$ model, the 
effective theory makes predictions that are exact, order by order in a
systematic low-energy expansion.

Hole-doped cuprates have hole pockets centered at lattice momenta 
$(\pm \frac{\pi}{2a},\pm \frac{\pi}{2a})$. The location of the hole pockets has
important consequences for the fermion field content of the effective
theory and on the realization of the various symmetries of these fields. 
In electron-doped cuprates the charged quasiparticles reside in momentum space
pockets centered at $(\frac{\pi}{a},0)$ or $(0,\frac{\pi}{a})$
\cite{Lee03,Kus03,Sen04,Yua04,Toh04,Guo06,Aic06}.
We have computed the single-electron dispersion relation in the $t$-$t'$-$J$
model shown in figure 1. The energy $E(\vec p)$ of an electron is indeed 
minimal when its lattice momentum $\vec p = (p_1,p_2)$ is located in an 
electron pocket centered at $(\frac{\pi}{a},0)$ or $(0,\frac{\pi}{a})$.
\begin{figure}[t]
\begin{center}
\vspace{-2.4cm}
\epsfig{file=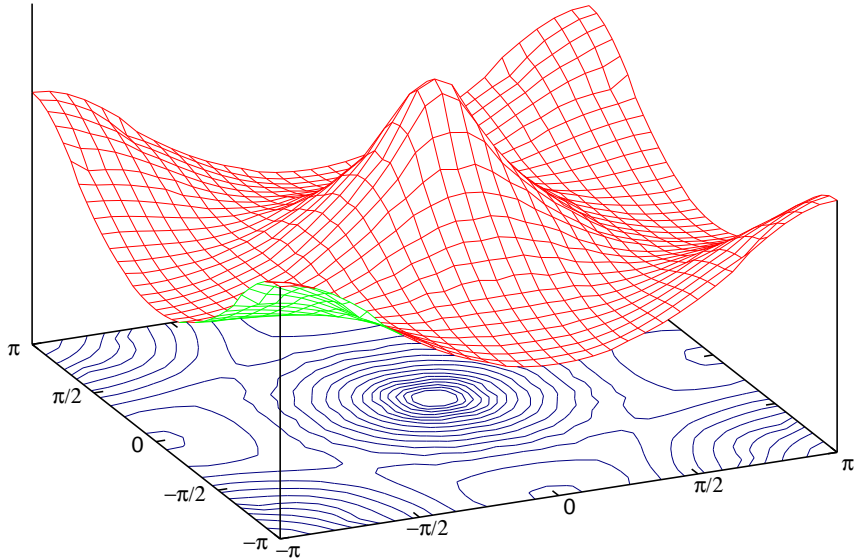,width=15cm}
\end{center}
\vspace{-1cm}
\caption{\it The dispersion relation $E(\vec p)$ of a single electron in the 
$t$-$t'$-$J$ model (on a $32 \times 32$ lattice for $J = 0.4 t$ and 
$t' =-0.3 t$) with electron pockets centered at  $(\frac{\pi}{a},0)$ and
$(0,\frac{\pi}{a})$.}
\end{figure}
The location of these pockets again has important effects on the electron 
dynamics, which turns out to be quite different from that of the holes. In 
particular, in contrast to hole-doped systems, in electron-doped 
antiferromagnets the 
magnon-mediated forces between two electrons depend on the total momentum
$\vec P$ of the pair. For $\vec P = 0$ the one-magnon exchange  potential 
between two electrons at distance $r$ is proportional to $1/r^4$, while in the 
hole case it has a $1/r^2$ dependence. The different locations of electron and 
hole pockets also affect the phase structure. While spiral phases are possible 
in the hole-doped case \cite{Shr88,Kan89,Bru06a}, they are absent in 
electron-doped cuprates \cite{Goo94,Yam99,Kus02,Yam03,Mar03}.

The paper is organized as follows. In section 2 the symmetries of charge 
carrier fields are summarized. Based on this, the electron fields are 
identified and the hole fields are eliminated. The low-energy effective action
for magnons and electrons is then constructed in a systematic manner. Section 3
contains the derivation of the one-magnon exchange potential between two 
electrons as well as a discussion of the corresponding Schr\"odinger equation. 
In section 4 spiral configurations of the staggered magnetization and in 
section 5 the reduction of the staggered magnetization upon doping are
investigated. Section 6 contains our conclusions. The somewhat subtle 
transformation of the one-magnon exchange potential from momentum to coordinate
space is discussed in an appendix.

\section{Symmetries of Magnon and Electron Fields}

In this section, based on \cite{Kae05,Bru06}, we summarize the transformation
properties of magnon and charge carrier fields. We then identify the electron 
fields and eliminate the hole fields in order to construct the low-energy 
effective theory for magnons and electrons.

\subsection{Symmetries of Magnon Fields}

In an antiferromagnet the spontaneous breaking of the $SU(2)_s$ spin symmetry 
down to $U(1)_s$ gives rise to two massless magnons. The staggered 
magnetization is described by a unit-vector field
\begin{equation}
\vec e(x) = (e_1(x),e_2(x),e_3(x)) = 
(\sin\theta(x) \cos\varphi(x),\sin\theta(x) \sin\varphi(x),\cos\theta(x)), 
\end{equation}
in the coset space $SU(2)_s/U(1)_s = S^2$, where $x = (x_1,x_2,t)$ is a point 
in $(2+1)$-d space-time. It is convenient to use a $\C P(1)$ representation in
terms of $2 \times 2$ Hermitean projection matrices $P(x)$ with
\begin{equation}
P(x) = \frac{1}{2}(\1 + \vec e(x) \cdot \vec \sigma), \quad
P(x)^\dagger = P(x), \quad \mbox{Tr} P(x) = 1, \quad P(x)^2 = P(x).
\end{equation}
As discussed in detail in \cite{Kae05}, the magnon field transforms as
\begin{alignat}{3}
SU(2)_s:&\quad &P(x)' &= g P(x) g^\dagger, \nonumber \\
SU(2)_Q:&\quad  &^{\vec Q}P(x) &= P(x), \nonumber \\
D_i:&\quad &^{D_i}P(x) &= \1 - P(x), \nonumber \\
D'_i:&\quad &^{D'_i}P(x) &= P(x)^*, \nonumber \\
O:&\quad &^OP(x) &= P(Ox), &\quad Ox &= (- x_2,x_1,t), \nonumber \\
R:&\quad &^RP(x) &= P(Rx), &\quad Rx &= (x_1,- x_2,t), \nonumber \\
T:&\quad &^TP(x) &= \1 - P(Tx), &\quad Tx &= (x_1,x_2,- t), \nonumber \\
T':&\quad &^{T'}P(x) &= (i \sigma_2) \left[^TP(x)\right] 
(i \sigma_2)^\dagger   = P(Tx)^*. \hspace{-4em}
\end{alignat}
The various symmetries are the $SU(2)_s$ spin rotations, the non-Abelian 
$SU(2)_Q$ extension of the $U(1)_Q$ fermion number symmetry (also known as 
pseudo-spin symmetry) that arises in the Hubbard model at half-filling, the 
displacement symmetry by one lattice spacing in the
$i$-direction $D_i$, the symmetry $D_i$ combined with the spin rotation
$i \sigma_2$ resulting in $D_i'$, as well as the 90 degrees rotation $O$, the
reflection at the $x_1$-axis $R$, time reversal $T$, and $T$ combined with the
spin rotation $i \sigma_2$ resulting in $T'$.

The spontaneously broken $SU(2)_s$ symmetry is nonlinearly realized on the
charge carrier fields. The global $SU(2)_s$ symmetry then manifests itself as a
local $U(1)_s$ symmetry in the unbroken subgroup, and the charge carrier fields
couple to the magnon field via composite vector fields. In order to construct
these vector fields one first diagonalizes $P(x)$ by a unitary transformation
$u(x) \in SU(2)$, i.e.
\begin{gather}
u(x) P(x) u(x)^\dagger = \frac{1}{2}(\1 + \sigma_3) = 
\left(\begin{array}{cc} 1 & 0 \\ 0 & 0 \end{array} \right), \qquad 
u_{11}(x) \geq 0, \nonumber \\[0.7ex]
u(x) = \left(\begin{array}{cc} \cos\frac{\theta(x)}{2} & 
\sin\frac{\theta(x)}{2} \exp(- i \varphi(x)) \\
- \sin\frac{\theta(x)}{2} \exp(i \varphi(x)) & \cos\frac{\theta(x)}{2} 
\end{array}\right).
\end{gather}
Under a global $SU(2)_s$ transformation $g$, the diagonalizing field $u(x)$
transforms as
\begin{equation}
\label{trafou}
u(x)' = h(x) u(x) g^\dagger, \qquad u_{11}(x)' \geq 0.
\end{equation}
This defines the nonlinear symmetry transformation 
\begin{equation}
h(x) = \exp(i \alpha(x) \sigma_3)
  = \left(\begin{array}{cc}
  \exp(i \alpha(x)) & 0 \\ 0 & \exp(- i \alpha(x)) \end{array} \right)
  \in U(1)_s.
\end{equation}
Under the displacement symmetry $D_i$ the staggered magnetization changes 
sign, i.e.\ $^{D_i}\vec e(x) = - \vec e(x)$, and one obtains 
\begin{equation}
^{D_i}u(x) = \tau(x) u(x), \qquad
\tau(x) = \left(\begin{array}{cc} 0 & - \exp(- i \varphi(x)) \\
\exp(i \varphi(x)) & 0 \end{array} \right).
\end{equation}
In order to couple magnons and charge carriers, one constructs the traceless 
anti-Hermitean field
\begin{equation}
v_\mu(x) = u(x) \p_\mu u(x)^\dagger,
\end{equation}
which transforms as
\begin{alignat}{3}
SU(2)_s:&\quad &v_\mu(x)' &= h(x) [v_\mu(x) + \p_\mu] h(x)^\dagger,
  \hspace{-5em} \nonumber \\[-.2ex]
SU(2)_Q:&\quad &^{\vec Q}v_\mu(x) &= v_\mu(x), \nonumber \\
D_i:&\quad &^{D_i}v_\mu(x) &= \tau(x)[v_\mu(x) + \p_\mu] \tau(x)^\dagger,
  \hspace{-5em} \nonumber \\
D'_i:&\quad &^{D'_i}v_\mu(x) &= v_\mu(x)^*, \nonumber \\
O:&\quad &^Ov_i(x) &= \varepsilon_{ij} v_j(Ox), \quad
  &^Ov_t(x) &= v_t(Ox), \nonumber \\
R:&\quad &^Rv_1(x) &= v_1(Rx), \quad &^Rv_2(x) &= - v_2(Rx),
  \quad ^Rv_t(x) = v_t(Rx), \nonumber \\
T:&\quad &^Tv_j(x) &= \ ^{D_i}v_j(Tx), \quad &^Tv_t(x) &= - \ ^{D_i}v_t(Tx), 
  \nonumber \\
T':&\quad &^{T'}v_j(x) &= \ ^{D'_i}v_j(Tx), \quad
   &^{T'}v_t(x) &= - ^{D'_i}v_t(Tx).
\end{alignat}
The field $v_\mu(x)$ decomposes into an Abelian ``gauge'' field $v_\mu^3(x)$ 
and two ``charged'' vector fields $v_\mu^\pm(x)$, i.e.
\begin{equation}
v_\mu(x) = i v_\mu^a(x) \sigma_a, \qquad
v_\mu^\pm(x) = v_\mu^1(x) \mp i v_\mu^2(x).
\end{equation}

\subsection{Fermion Fields in Momentum Space Pockets}

In \cite{Bru06} matrix-valued charge carrier fields
\begin{equation}
\Psi^k(x) = \left(\begin{array}{cc} \psi^k_+(x) & \psi^{-k'\dagger}_-(x) 
\\ \psi^k_-(x) & - \psi^{-k' \dagger}_+(x) \end{array} \right), \quad
\Psi^{k\dagger}(x) = \left(\begin{array}{cc} 
\psi^{k\dagger}_+(x) & \psi^{k\dagger}_-(x) \\ 
\psi^{-k'}_-(x) & - \psi^{-k'}_+(x) \end{array} \right)
\end{equation}
have been constructed. Here $k' = k + \big(\frac{\pi}{a},\frac{\pi}{a}\big)$
and $\psi^k_\pm(x)$ and $\psi^{k\dagger}_\pm(x)$ are independent Grassmann
fields which are associated with the following eight lattice momentum values 
illustrated in figure 2
\begin{equation}
k = (k_1,k_2) \in 
\left\{\big(0,0\big),\; \big(\frac{\pi}{a},\frac{\pi}{a}\big),\;
\big(\frac{\pi}{a},0\big),\; \big(0,\frac{\pi}{a}\big),\;
\big(\pm \frac{\pi}{2a},\pm \frac{\pi}{2a}\big)\right\}.
\end{equation}
\begin{figure}[t]
\begin{center}
\vspace{-0.4cm}
\epsfig{file=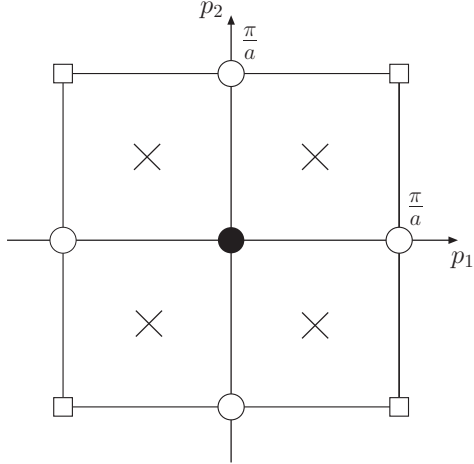,width=7cm}
\end{center}
\caption{\it Eight lattice momenta and their periodic copies. In the cuprates 
the holes reside in momentum space pockets centered at lattice momenta 
$\big(\pm \frac{\pi}{2a},\pm \frac{\pi}{2a}\big)$ which are represented by the 
four crosses, while electrons reside at $\big(\frac{\pi}{a},0\big)$ or 
$\big(0,\frac{\pi}{a}\big)$ (represented by the circles).}
\end{figure}
The charge carrier fields transform as
{\allowdisplaybreaks
\begin{alignat}{2}
\label{phitrafo}
SU(2)_s:&\quad &\Psi^k(x)' &= h(x) \Psi^k(x), \nonumber \\[-.2ex]
SU(2)_Q:&\quad &^{\vec Q}\Psi^k(x) &= \Psi^k(x) \Omega^T, \nonumber \\
D_i:&\quad &^{D_i}\Psi^k(x) &= \exp(i k_i a) \tau(x) \Psi^k(x) \sigma_3, 
\nonumber \\
D'_i:&\quad &^{D'_i}\Psi^k(x) &= 
\exp(i k_i a) (i \sigma_2) \Psi^k(x) \sigma_3, \nonumber \\\pagebreak
O:&\quad &^O\Psi^k(x) &= \Psi^{Ok}(Ox), \nonumber \\
R:&\quad &^R\Psi^k(x) &= \Psi^{Rk}(Rx), \nonumber \\
T:&\quad &^T\Psi^k(x) &= 
\tau(Tx) (i \sigma_2) \left[\Psi^{-k\dagger}(Tx)^T\right] \sigma_3, 
\nonumber \\
&\quad &^T\Psi^{k\dagger}(x) &= - \sigma_3 \left[\Psi^{-k}(Tx)^T\right] 
(i \sigma_2)^\dagger \tau(Tx)^\dagger, \nonumber \\
T':&\quad &^{T'}\Psi^k(x) &= - \left[\Psi^{-k\dagger}(Tx)^T\right] \sigma_3, 
\nonumber \\
&\quad &^{T'}\Psi^{k\dagger}(x) &= \sigma_3 \left[\Psi^{-k}(Tx)^T\right].
\end{alignat}
} 
Here $\Omega \in SU(2)_Q$ and $Ok$ and $Rk$ are the momenta obtained by 
rotating or reflecting the momentum $k$.

\subsection{Electron Field Identification and Hole Field Elimination}

ARPES measurements as well as theoretical investigations
\cite{Lee03,Kus03,Sen04,Yua04,Toh04,Guo06,Aic06} (see also figure 1) indicate 
that electrons doped into an antiferromagnet appear in momentum space pockets 
centered at 
\begin{equation}
k = \big(\frac{\pi}{a},0\big), \qquad k' = \big(0,\frac{\pi}{a}\big). 
\end{equation}
Hence, only the fermion fields with these two momentum labels will appear in 
the low-energy effective theory. Using the transformation rules of 
eq.(\ref{phitrafo}) one can construct the following invariant mass terms
\begin{align} 
\frac{1}{2}
\mbox{Tr} \big[ {\cal M} (\Psi^{k\dagger} \sigma_3 & \Psi^{k'} +
\Psi^{k'\dagger} \sigma_3 \Psi^k) + m (\Psi^{k\dagger} \Psi^k \sigma_3 +
\Psi^{k'\dagger} \Psi^{k'} \sigma_3)\big] \nonumber \\
 = & \; \; \;
{\cal M} \big(\psi^{k\dagger}_+ \psi^{k'}_+ - \psi^{k\dagger}_- \psi^{k'}_- +
\psi^{k'\dagger}_+ \psi^k_+ - \psi^{k'\dagger}_- \psi^k_- \big) 
\nonumber \\
& + m \big(\psi^{k\dagger}_+ \psi^k_+ + 
\psi^{k\dagger}_- \psi^k_- + \psi^{k'\dagger}_+ \psi^{k'}_+ + 
\psi^{k'\dagger}_- \psi^{k'}_- \big) \nonumber \\[0.3ex]
= & \; \; \;
\big(\psi^{k\dagger}_+, \, \psi^{k'\dagger}_+ \big) 
\bigg(\begin{array}{cc} m & {\cal M} \\ {\cal M} & m \end{array}\bigg)
\bigg(\begin{array}{c} \psi^k_+ \\ \psi^{k'}_+ \end{array}\bigg) 
\nonumber \\
& + \big(\psi^{k\dagger}_-, \, \psi^{k'\dagger}_- \big)
\bigg(\begin{array}{cc} m & - {\cal M} \\ - {\cal M} & m \end{array}\bigg)
\bigg(\begin{array}{c} \psi^k_- \\ \psi^{k'}_- \end{array}\bigg).
\end{align}
The terms proportional to ${\cal M}$ are $SU(2)_Q$-invariant while those
proportional to $m$ are only $U(1)_Q$-invariant. By diagonalizing the mass 
matrices, electron and hole fields can be identified. The resulting eigenvalues
are $m \pm {\cal M}$. In the $SU(2)_Q$-symmetric case, i.e.\ for $m = 0$, there
is an electron-hole symmetry. The electrons correspond to 
positive energy states with eigenvalue ${\cal M}$ and the holes correspond to 
negative energy states with eigenvalue $- {\cal M}$. In the presence of 
$SU(2)_Q$-breaking terms these energies are shifted and electrons now
correspond to states with eigenvalue $m + {\cal M}$, while holes correspond to 
states with eigenvalue $m - {\cal M}$. The electron fields are given by the 
corresponding eigenvectors
\begin{equation}
\psi_+(x) = \frac{1}{\sqrt{2}}
\big[ \psi^k_+(x) + \psi^{k'}_+(x) \big], \qquad
\psi_-(x) = \frac{1}{\sqrt{2}}
\big[ \psi^k_-(x) - \psi^{k'}_-(x) \big].
\end{equation}
Under the various symmetries they transform as
\begin{alignat}{2}
\label{symcomp}
SU(2)_s:&\quad &\psi_\pm(x)' &= \exp(\pm i \alpha(x)) \psi_\pm(x),
\nonumber \\
U(1)_Q:&\quad &^Q\psi_\pm(x) &= \exp(i \omega) \psi_\pm(x),
\nonumber \\
D_i:&\quad &^{D_i}\psi_\pm(x) &= 
\mp \exp(i k_i a) \exp(\mp i \varphi(x)) \psi_\mp(x),
\nonumber \\
D'_i:&\quad &^{D'_i}\psi_\pm(x) &= \pm \exp(i k_i a) \psi_\mp(x),
\nonumber \\
O:&\quad &^O\psi_\pm(x) &= \pm \psi_\pm(Ox), \nonumber \\
R:&\quad &^R\psi_\pm(x) &= \psi_\pm(Rx), \nonumber \\
T:&\quad &^T\psi_\pm(x) &= \exp(\mp i \varphi(Tx)) 
\psi^\dagger_\pm(Tx),
\nonumber \\
&\quad &^T\psi^\dagger_\pm(x) &= - \exp(\pm i \varphi(Tx)) \psi_\pm(Tx),
\nonumber \\
T':&\quad &^{T'}\psi_\pm(x) &= - \psi^\dagger_\pm(Tx), \nonumber \\
&\quad &^{T'}\psi^\dagger_\pm(x) &= \psi_\pm(Tx).
\end{alignat}
The action of magnons and electrons must be invariant under these symmetries.

\subsection{Effective Action for Magnons and Electrons}

We decompose the action into terms containing different numbers of fermion 
fields $n_\psi$ (with $n_\psi$ even) such that
\begin{equation}
S[\psi^\dagger_\pm,\psi_\pm,P] = \int d^2x \ dt \ \sum_{n_\psi}
{\cal L}_{n_\psi}.
\end{equation}
The leading terms in the effective Lagrangian without fermion fields are given
by
\begin{equation}
{\cal L}_0 = 
\rho_s \mbox{Tr} \big[ \p_i P \p_i P + \frac{1}{c^2} \p_t P \p_t P \big],
\end{equation}
with the spin stiffness $\rho_s$ and the spinwave velocity $c$ as low-energy
parameters. The terms with two fermion fields (containing at most one temporal 
or two spatial derivatives) describe the propagation of electrons as well as 
their couplings to magnons, and are given by
\begin{align}
\label{action}
{\cal L}_2 = & \sum_{s = +,-} \Big[
M \psi^\dagger_s \psi_s + \psi^\dagger_s D_t \psi_s +
\frac{1}{2 M'} D_i \psi^\dagger_s D_i \psi_s  + 
N \psi^\dagger_s v^s_i v^{-s}_i \psi_s \nonumber \\
& \hspace{1.8em}  + i K \big(D_1 \psi^\dagger_s v^s_1 \psi_{-s} -
\psi^\dagger_s v^s_1 D_1 \psi_{-s} -
D_2 \psi^\dagger_s v^s_2 \psi_{-s} +
\psi^\dagger_s v^s_2 D_2 \psi_{-s}\big)\Big].
\end{align}
Here $M$ is the rest mass and $M'$ is the kinetic mass of an electron, $K$ is
an electron-one-magnon, and $N$ is an electron-two-magnon coupling, which all
take real values. The covariant derivatives are given by
\begin{align}
D_t \psi_\pm(x) &= \left[\p_t \pm i v_t^3(x) - \mu \right] \psi_\pm(x),
\nonumber \\
D_i \psi_\pm(x) &= \left[\p_i \pm i v_i^3(x)\right] \psi_\pm(x).
\end{align}
The chemical potential $\mu$ enters the covariant time-derivative like an
imaginary constant vector potential for the fermion number symmetry $U(1)_Q$.

Next we list the contributions with four fermion fields including up to one 
temporal or two spatial derivatives
\begin{align}
\label{Lagrange}
{\cal L}_4 = & \sum_{s = +,-} \Big[
\frac{G_1}{2} \psi^\dagger_s \psi_s \psi^\dagger_{-s} \psi_{-s} +
G_2 D_i \psi^\dagger_s D_i \psi_s \psi^\dagger_s \psi_s +
G_3 D_i \psi^\dagger_s D_i \psi_s \psi^\dagger_{-s} \psi_{-s} \nonumber 
\\[-1.5ex]
&\hspace{1.9em} +G_4 D_i \psi^\dagger_s D_i\psi_{-s} \psi^\dagger_{-s} \psi_s +
\frac{G_5}{2} \big( D_i \psi^\dagger_s \psi_s D_i \psi^\dagger_{-s} \psi_{-s} +
\psi^\dagger_s D_i \psi_s \psi^\dagger_{-s} D_i \psi_{-s} \big)
\nonumber \\[0.2ex]
&\hspace{1.9em} +i G_6 \big( D_1 \psi^\dagger_s \psi_s \psi^\dagger_s v^s_1
  \psi_{-s}  - \psi^\dagger_s D_1 \psi_s \psi^\dagger_{-s} v^{-s}_1 \psi_s
\nonumber \\
&\hspace{3.8em} - D_2 \psi^\dagger_s \psi_s \psi^\dagger_s v^s_2 \psi_{-s}
+\psi^\dagger_s D_2 \psi_s \psi^\dagger_{-s} v^{-s}_2 \psi_s \big) +
\frac{G_7}{2} \psi^\dagger_s\psi_s v^s_i v^{-s}_i \psi^\dagger_{-s}\psi_{-s}
\nonumber \\
&\hspace{1.9em} + \frac{G_8}{2} \big( D_t \psi^\dagger_s \psi_s 
\psi^\dagger_{-s}
  \psi_{-s}  -\psi^\dagger_s D_t \psi_s \psi^\dagger_{-s} \psi_{-s} \big)\Big].
\end{align}
Since it contains $D_t$, the term proportional to $G_8$ would imply a deviation
from canonical anticommutation relations in a Hamiltonian formulation of the
theory. Fortunately, this term can be eliminated by a field redefinition
$\psi_s \rightarrow \psi_s + \frac{G_8}{2} \psi_s \psi_{-s}^\dagger \psi_{-s}$.
The redefined field obeys the same symmetry transformations as the original 
one and is constructed such that after the field redefinition $G_8 = 0$. All
other terms in the action are reproduced in their present form.

For completeness, we finally list the only contribution with more than four
fermion fields, again including up to one temporal or two spatial derivatives
\begin{align}
{\cal L}_6 = \sum_{s = +,-} &
H D_i \psi^\dagger_s D_i \psi_s \psi^\dagger_s \psi_s \psi^\dagger_{-s}
\psi_{-s}.
\end{align}
The leading fermion contact term is proportional to $G_1$. Due to the large 
number of low-energy parameters $G_2,...,G_7,H$, the higher-order terms are 
unlikely to be used in practical applications. We have used the algebraic 
program FORM \cite{Ver00}, and independently thereof, the GiNaC framework for
symbolic computation within the C++ programming language \cite{Bau02}, to
verify that the terms listed above form a complete linearly independent set.

It should be noted that, unlike in the hole case, the leading terms in the
effective action are not invariant against Galilean boosts. This is not 
unexpected because the underlying microscopic systems also lack this symmetry.
The lack of Galilean boost invariance has important physical consequences. In
particular, the magnon-mediated forces between two electrons will turn out to
depend on the total momentum $\vec P$ of the pair. Thus it is not sufficient to
consider the two particles in their rest frame, i.e.\ at $\vec P = 0$. This is 
due to the underlying crystal lattice which defines a preferred rest-frame (a 
condensed matter ``ether'').

\section{Magnon-mediated Binding between Electrons}

We treat the forces between two electrons in the same way as the ones in 
the effective theory for magnons and holes \cite{Bru05,Bru06}. As in that case,
one-magnon exchange dominates the long-range forces. In this section we 
calculate the one-magnon exchange potential between two electrons and we solve 
the corresponding two-particle Schr\"odinger equation.

\subsection{One-Magnon Exchange Potential between Electrons}

In order to calculate the one-magnon exchange potential between two electrons,
we expand in the magnon fluctuations $m_1(x)$, $m_2(x)$ around the ordered 
staggered magnetization, i.e.
\begin{align}
\vec e(x) = \Big( \frac{m_1(x)}{\sqrt{\rho_s}},\,&
\frac{m_2(x)}{\sqrt{\rho_s}},1 \Big) + {\cal O}(m^2) \nonumber \\
\Rightarrow \quad
v_\mu^\pm(x) &= \frac{1}{2 \sqrt{\rho_s}} \p_\mu
\big[ m_2(x) \pm i m_1(x) \big] + {\cal O}(m^3), \nonumber \\
v_\mu^3(x) &= \frac{1}{4 \rho_s}\big[m_1(x) \p_\mu m_2(x) -
m_2(x) \p_\mu m_1(x)\big] + {\cal O}(m^4).
\end{align}
The vertices with $v_\mu^3(x)$ (contained in $D_\mu$) involve at least two 
magnon fields. Hence, one-magnon exchange results exclusively from vertices 
with $v_\mu^\pm(x)$. Thus, two electrons can exchange a single magnon only if 
they have antiparallel spins ($+$ and $-$), which are both flipped in the 
magnon exchange process. We denote the momenta of the incoming and outgoing 
electrons by $\vec p_\pm$ and $\vec p_\pm \ \!\!\!\! ' \ $, respectively. 
Furthermore, $\vec q$ represents the momentum of the exchanged magnon. We also
introduce the total momentum $\vec P$ as well as the incoming and outgoing 
relative momenta 
$\vec p$ and $\vec p \ '$
\begin{gather}
\vec P = \vec p_+ + \vec p_- = \vec p_+ \ \!\!\!\! ' + \vec p_- \ \!\!\!\! '\:,
\nonumber \\
\vec p = \frac{1}{2}(\vec p_+ - \vec p_-), \qquad
\vec p \ ' = \frac{1}{2}(\vec p_+ \ \!\!\!\! ' - \vec p_- \ \!\!\!\! ').
\end{gather}
Due to momentum conservation we then have
\begin{equation}
\vec q = \vec p + \vec p \ '.
\end{equation}
Figure 3 shows the Feynman diagram describing one-magnon exchange.
\begin{figure}[tb]
\begin{center}
\vspace{-0.4cm}
\epsfig{file=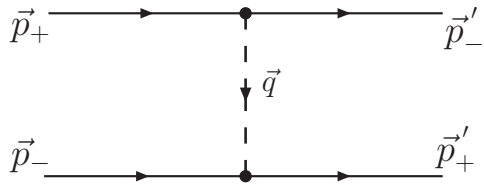,width=7cm}
\end{center}
\caption{\it Feynman diagram for one-magnon exchange between two electrons with
antiparallel spins undergoing a spin-flip.}
\end{figure}
In momentum space the resulting one-magnon exchange potential takes the form
\begin{align}
\langle \vec p_+ \ \!\!\!\! ' \ \vec p_- \ \!\!\!\! '|V
|\vec p_+ \vec p_-\rangle &= \frac{K^2}{2 \rho_s} \frac{1}{q^2}
\left[q_1^2 - q_2^2 + 2(q_1 p_{-1} - q_2 p_{-2})\right]
\left[q_1^2 - q_2^2 - 2(q_1 p_{+1} - q_2 p_{+2})\right] \nonumber \\
&\;\;\;\; \times
\delta(\vec p_+ + \vec p_- - \vec p_+ \ \!\!\!\! ' - \vec p_- \ \!\!\!\! ').
\end{align}
Transforming the potential to coordinate space is not entirely trivial and is
thus discussed in the appendix. In coordinate space the resulting potential is 
given by
\begin{equation}
\langle \vec r_+ \ \!\!\!\! ' \vec r_- \ \!\!\!\! '|V
|\vec r_+ \vec r_-\rangle = \frac{K^2}{2 \pi \rho_s} 
\left[12 \frac{\cos(4 \varphi)}{r^4} + 
\frac{P^2}{2} \frac{\cos(2 (\varphi + \chi))}{r^2}
\right] \delta(\vec r_+ - \vec r_- \ \!\!\!\! ') \
\delta(\vec r_- - \vec r_+ \ \!\!\!\! ').
\end{equation}
Here $\varphi$ is the angle between the distance vector 
$\vec r = \vec r_+ - \vec r_-$ of the two electrons and the $x_1$-axis. In
contrast to the hole case, the potential depends on the magnitude $P$ of the
total momentum $\vec P$ as well as on the angle $\chi$ between $\vec P$ and
the $x_1$-axis. For $\vec P = 0$ the one-magnon exchange potential between two
electrons falls off as $1/r^4$, while in the hole case it is proportional to
$1/r^2$. Retardation effects enter at higher orders only and thus the potential
is instantaneous. We have omitted short-distance $\delta$-function 
contributions to the potential which add to the 4-fermion contact interactions.
Since we will model the short-distance repulsion by a hard-core radius, the 
$\delta$-function contributions will not be needed in the following.

\subsection{Schr\"odinger Equation for two Electrons}

Let us consider two electrons with opposite spins $+$ and $-$. The wave 
function depends on the relative distance vector $\vec r$ which points from the
spin $-$ electron to the spin $+$ electron. Magnon exchange is accompanied by a
spin-flip. Hence, the vector $\vec r$ changes its direction in the magnon 
exchange process. The resulting Schr\"odinger equation then takes the form
\begin{equation}
- \frac{1}{M'} \Delta \Psi(\vec r) + \frac{K^2}{2 \pi \rho_s} 
\left[12 \frac{\cos(4 \varphi)}{r^4} + 
\frac{P^2}{2} \frac{\cos(2 (\varphi + \chi))}{r^2}
\right] \Psi(- \vec r) = \bigg[ E - \frac{P^2}{2 M'} \bigg] \Psi(\vec r).
\end{equation}
For simplicity, instead of explicitly using the 4-fermion contact interactions,
we model the short-distance repulsion between the electrons by a hard-core of 
radius $r_0$, i.e.\ we require $\Psi(\vec r) = 0$ for $|\vec r| \leq r_0$.
In contrast to the hole case \cite{Bru05,Bru06}, we have not been able to solve
the above Schr\"odinger equation analytically. Instead, we have solved it
numerically. A typical probability distribution for the ground state is
illustrated in figure 4 for $\vec P = 0$.
\begin{figure}[tb]
\begin{center}
\vspace{-0.3cm}
\epsfig{file=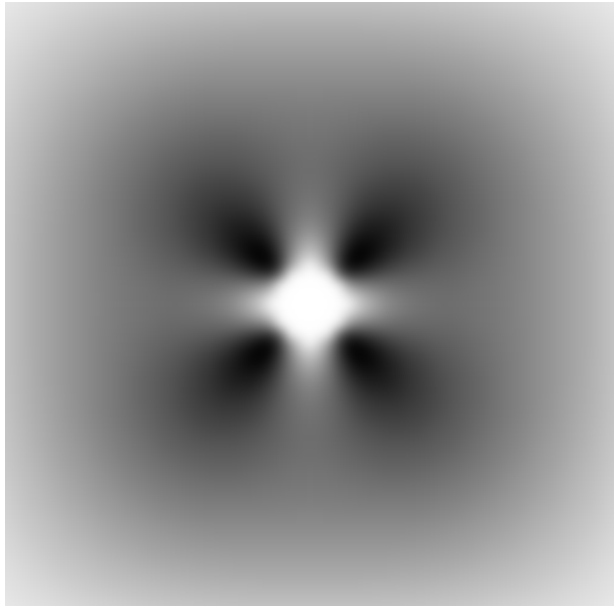,width=8cm}
\end{center}
\caption{\it Probability distribution for the ground state of two electrons
with total momentum $\vec P = (0,0)$.}
\end{figure}
The probability distribution resembles $d_{xy}$ symmetry. However, due to the 
90 degrees rotation symmetry the continuum classification scheme of angular 
momenta is inappropriate. Under the group of discrete rotations and reflections
the ground state wave function transforms in the trivial representation.

Due to the lack of Galilean boost invariance, the two-electron bound state
changes its structure when it is boosted out of its rest frame. Of course, an
electron pair with total momentum $\vec P \neq 0$ costs additional kinetic
energy $P^2/2 M'$ for the center of mass motion. In addition, the binding
energy also depends on $\vec P$. The strongest binding arises when the total
momentum $\vec P$ points along a lattice diagonal. The corresponding 
probability distribution is illustrated in figure 5.
\begin{figure}[tb]
\begin{center}
\vspace{-0.3cm}
\epsfig{file=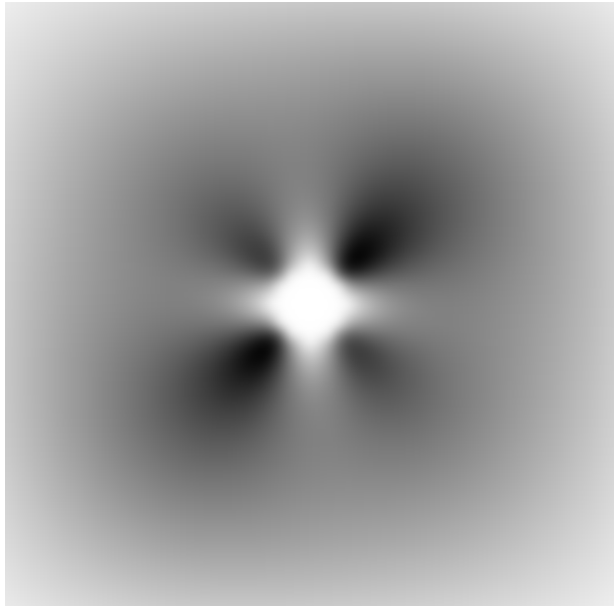,width=8cm}
\end{center}
\caption{\it Probability distribution for the ground state of two electrons
with total momentum $\vec P = \frac{1}{\sqrt{2}}(P,P)$ along a lattice 
diagonal.}
\end{figure}
Since they depend crucially on the precise values of the low-energy parameters,
we have not attempted an extensive numerical investigation of the binding 
energy and other properties of the two-electron bound states. Once the 
low-energy parameters have been determined for a concrete underlying 
microscopic system, a precise calculation of the physical properties of the
two-electron bound state is straightforward using the numerical method employed
above.

In order to gain at least some approximate analytic insight into the bound 
state problem, let us also consider the semi-classical Bohr-Sommerfeld 
quantization. First, we consider a pair of electrons with total momentum 
$\vec P = 0$ moving relative to each other along a lattice diagonal. The 
classical energy of the periodic relative motion is given by
\begin{equation}
E = M' \left(\frac{dr}{dt}\right)^2 - \frac{6 K^2}{\pi \rho_s r^4}.
\end{equation}
The Bohr-Sommerfeld quantization condition implies
\begin{eqnarray}
S + E T&=&\int_0^T dt \ \left[M' \left(\frac{dr}{dt}\right)^2 +
\frac{6 K^2}{\pi \rho_s r^4} + E\right] =
2 \int_0^T dt \ M' \left(\frac{dr}{dt}\right)^2 \nonumber \\
&=&4 \int_{r_0}^R dr \ M' \ \frac{dr}{dt} =
4 \int_{r_0}^R dr \ \sqrt{E M' + \frac{6 K^2 M'}{\pi \rho_s r^4}} = 2 \pi n,
\end{eqnarray}
where $S$ is the action, $T$ is the period of the motion, and $n$ is a positive
integer. The hard-core radius $r_0$ is a classical turning point and $R$ is the
other classical turning point determined by
\begin{equation}
E = - \frac{6 K^2}{\pi \rho_s R^4}.
\end{equation} 
The above equations lead to a relatively complicated expression for the energy 
in terms of elliptic integrals. Instead of investigating these expressions, we 
limit ourselves to estimating the number of bound states. For this purpose, we 
set $E = 0$ which implies $R =\infty$, and we then obtain
\begin{equation}
n = \left[ \ \int_{r_0}^\infty dr \ \sqrt{\frac{24 K^2 M'}{\pi^3 \rho_s r^4}}
\ \right] = \left[ \ \sqrt{\frac{24 K^2 M'}{\pi^3 \rho_s r_0^2}} \ \right].
\end{equation}
The brackets denote the nearest integer smaller than the expression enclosed in
the brackets. In particular, Bohr-Sommerfeld quantization suggests that a bound
state exists only if
\begin{equation}
\frac{24 K^2 M'}{\pi^3 \rho_s r_0^2} \geq 1.
\end{equation}
Of course, one should be aware of the fact that this is at best a 
semi-quantitative estimate because Bohr-Sommerfeld quantization should not be 
trusted quantitatively for small quantum numbers. Let us also repeat these 
considerations for $\vec P \neq 0$. Again, we consider $\vec P = 
\frac{1}{\sqrt{2}} (P,P)$ such that the diagonal motion of an electron-pair 
has the energy
\begin{equation}
E = M' \left(\frac{dr}{dt}\right)^2 - \frac{K^2}{2 \pi \rho_s} 
\left(\frac{12}{r^4} + \frac{P^2}{2 r^2}\right).
\end{equation}
In complete analogy to the $\vec P = 0$ case one then obtains
\begin{equation}
n = \left[ \ \int_{r_0}^\infty dr \ 
\sqrt{\frac{2 K^2 M'}{\pi^3 \rho_s} \left(\frac{12}{r^4}
+ \frac{P^2}{2 r^2}\right)} \ \right] \rightarrow \infty,
\end{equation}
which suggests that infinitely many two-electron bound states exist for
$\vec P \neq 0$. This is similar to the two-hole problem which has a $1/r^2$
potential with infinitely many bound states already for $\vec P = 0$ 
\cite{Bru05,Bru06}.

Two-electron bound states with $\vec P = 0$ have been considered before by 
Kuchiev and Sushkov \cite{Kuc95} in the context of the $t$-$t'$-$J$ model. In 
contrast to the hole-case \cite{Kuc93} with a $1/r^2$ potential and an infinite
number of bound states, in the electron-case only a finite number of bound 
states was found. While some results of that work agree qualitatively with the
results of our effective theory, there are also important differences. For
example, in \cite{Kuc95} the magnon-electron vertex was considered to be the 
same as the magnon-hole vertex, while the two vertices are different in the 
effective theory.

\section{Investigation of Spiral Phases}

In the following we will investigate phases with constant fermion density. The
most general magnon field configuration $\vec e(x)$ which provides a constant 
background field for the doped electrons is not necessarily constant itself, 
but may represent a spiral in the staggered magnetization. While a spiral costs
magnetic energy proportional to the spin stiffness $\rho_s$, the electrons
might lower their energy by propagating in the spiral background. However, we
will find that spiral phases are not energetically favorable in electron-doped
systems.

\subsection{Spirals with Uniform Composite Vector Fields}

Since the electrons couple to the composite vector field $v_i(x)$ in a gauge 
covariant way, in order to provide a constant background field for the 
electrons, $v_i(x)$ must be constant up to a gauge transformation, i.e.
\begin{align}
\label{const}
{v^3_i}(x)'&=v^3_i(x) - \p_i \alpha(x) = 
\p_i \varphi(x) \sin^2\frac{\theta(x)}{2} - \p_i \alpha(x) = c^3_i,
\nonumber \\[.5ex]
{v^\pm_i}(x)'&=v^\pm_i(x) \exp\big(\pm 2 i \alpha(x)\big) \nonumber \\
&=\frac{1}{2} \big[\p_i \varphi(x) \sin\theta(x) 
\pm i \p_i \theta(x)\big] 
\exp\big(\pm 2 i \alpha(x) \mp i \varphi(x)\big) = c^\pm_i,
\end{align}
with $c^3_i$ and $c^\pm_i$ being constant. As shown in \cite{Bru06a}, the most
general configuration that leads to a constant $v_i(x)'$ represents a spiral in
the staggered magnetization. In addition, by an appropriate gauge 
transformation one can always put
\begin{equation}
c_i^+ = c_i^- = c_i \in \R.
\end{equation}
The magnetic energy density of such configurations takes the form
\begin{equation}
\epsilon_m = \frac{\rho_s}{2} \p_i \vec e(x) \cdot \p_i \vec e(x) = 
2 \rho_s v_i^+(x) v_i^-(x) = 2 \rho_s  c_i c_i.
\end{equation}

We now consider a concrete family of spiral configurations with
\begin{equation}
\theta(x) = \theta_0, \qquad \quad \varphi(x) = k_i x_i,
\end{equation}
which implies
\begin{equation}
v_t(x) = 0, \qquad v^3_i(x) = k_i \sin^2\frac{\theta_0}{2}, \qquad
v_i^\pm(x) = \frac{k_i}{2} \sin\theta_0 \exp(\mp i k_i x_i).
\end{equation}
Performing the gauge transformation
\begin{equation}
\alpha(x) = \frac{1}{2} k_i x_i,
\end{equation}
one arrives at
\begin{align}
\label{constant}
v_t(x)' &= v_t(x) - \p_t \alpha(x) = 0, \nonumber \\[.2ex]
{v^3_i}(x)' &= v^3_i(x) - \p_i \alpha(x) =
k_i \bigg[\sin^2\frac{\theta_0}{2} - \frac{1}{2}\bigg] = c^3_i,
\nonumber \\[-.4ex]
{v^\pm_i}(x)' &= v^\pm_i(x) \exp\big(\pm 2 i \alpha(x)\big) = 
\frac{k_i}{2} \sin\theta_0 = c_i,
\end{align}
such that
\begin{equation}
c^3_i = - \frac{k_i}{2} \cos\theta_0, \qquad \quad
a = \frac{|c_i|}{c^3_i} = - \tan\theta_0.
\end{equation}
The magnetic energy density then takes the form
\begin{equation}
\epsilon_m = 2 \rho_s c_i c_i =
\frac{\rho_s}{2} \big(k_1^2 + k_2^2\big) \sin^2\theta_0.
\end{equation}

\subsection{Fermionic Contributions to the Energy}

Let us now compute the fermionic contribution to the energy, first keeping
the parameters $c^3_i$ and $c_i$ of the spiral fixed, and neglecting the 
4-fermion contact interactions. The Euclidean action of eq.(\ref{action})
implies the following fermion Hamiltonian
\begin{align}
H = &\sum_{s = +,-} \Big[
M \Psi^\dagger_s \Psi_s + 
\frac{1}{2 M'} D_i \Psi^\dagger_s D_i \Psi_s +
N \Psi^\dagger_s v^s_i v^{-s}_i \Psi_s \nonumber \\
&\hspace{1.8em} + i K \big(D_1 \Psi^\dagger_s v^s_1 \Psi_{-s} -
\Psi^\dagger_s v^s_1 D_1 \Psi_{-s} -
D_2 \Psi^\dagger_s v^s_2 \Psi_{-s} +
\Psi^\dagger_s v^s_2 D_2 \Psi_{-s}\big)\Big],
\end{align}
with the covariant derivative
\begin{equation}
D_i \Psi_\pm(x) = [\p_i \pm i v^3_i(x)] \Psi_\pm(x).
\end{equation}
Here $\Psi^\dagger_\pm(x)$ and $\Psi_\pm(x)$ are creation and annihilation 
operators (not Grassmann numbers) for electrons with spin parallel ($+$) 
or antiparallel ($-$) to the local staggered magnetization. The Hamiltonian is 
invariant under time-independent $U(1)_s$ gauge transformations
\begin{align}
\Psi_\pm(x)' &= \exp\big(\pm i \alpha(x)\big) \Psi_\pm(x), \nonumber \\
{v^3_i}(x)' &= v^3_i(x) - \p_i \alpha(x), \nonumber \\
{v^\pm_i}(x)' &= v^\pm_i(x) \exp\big(\pm 2 i \alpha(x)\big).
\end{align}
We now consider electrons propagating in the background of a spiral in the 
staggered magnetization with
\begin{equation}
{v^3_i}(x)' = c^3_i, \qquad \quad {v^\pm_i}(x)' = c_i \in \R.
\end{equation}
After an appropriate gauge transformation, the fermions propagate in a constant
composite vector field ${v_i}(x)'$. The Hamiltonian is diagonalized by going to
momentum space. The Hamiltonian for single electrons with spatial momentum 
$\vec p = (p_1,p_2)$ is given by
\begin{equation}
H(\vec p) = \left(\begin{array}{cc}
M + \frac{(p_i - c_i^3)^2}{2 M'} + N c_i c_i &
2 K (- p_1 c_1 + p_2 c_2) \\ 2 K (- p_1 c_1 + p_2 c_2)  &
M + \frac{(p_i + c_i^3)^2}{2 M'} + N c_i c_i \end{array} \right).
\end{equation}
The diagonalization of the Hamiltonian yields
\begin{equation}
\label{energy}
E_\pm(\vec p) = M + \frac{p_i^2 + (c_i^3)^2}{2 M'} + N c_i c_i \pm 
\sqrt{\bigg(\frac{p_i c_i^3}{M'}\bigg)^2 + 4 K^2(p_1 c_1 - p_2 c_2)^2}.
\end{equation}
It should be noted that in this case the index $\pm$ does not refer to the spin
orientation. In fact, the eigenvectors corresponding to $E_\pm(\vec p)$ are 
linear combinations of both spins. Since $c_i^3$ does not affect the magnetic 
contribution to the energy density, it
can be fixed by minimizing $E_\pm(\vec p)$ which leads to $c_1^3 = c_2^3 = 0$. 
According to eq.(\ref{constant}) this implies that $\theta_0 = \frac{\pi}{2}$,
i.e.\ the spiral is along a great circle on the sphere $S^2$. For $c_i^3 = 0$ 
the energies of eq.(\ref{energy}) reduce to
\begin{equation}
E_\pm(\vec p) = M + \frac{p_i^2}{2 M'} + N c_i c_i 
\pm 2 K |p_1 c_1 - p_2 c_2|.
\end{equation}
The lines of constant energy are shown in figure 6. In particular, the lines of
constant $E_-(\vec p)$ are circles centered around $\pm 2 K M' (c_1,- c_2)$. 
\begin{figure}[t]
\begin{center}
\vspace{-0.4cm}
\begin{tabular}{cc} \epsfig{file=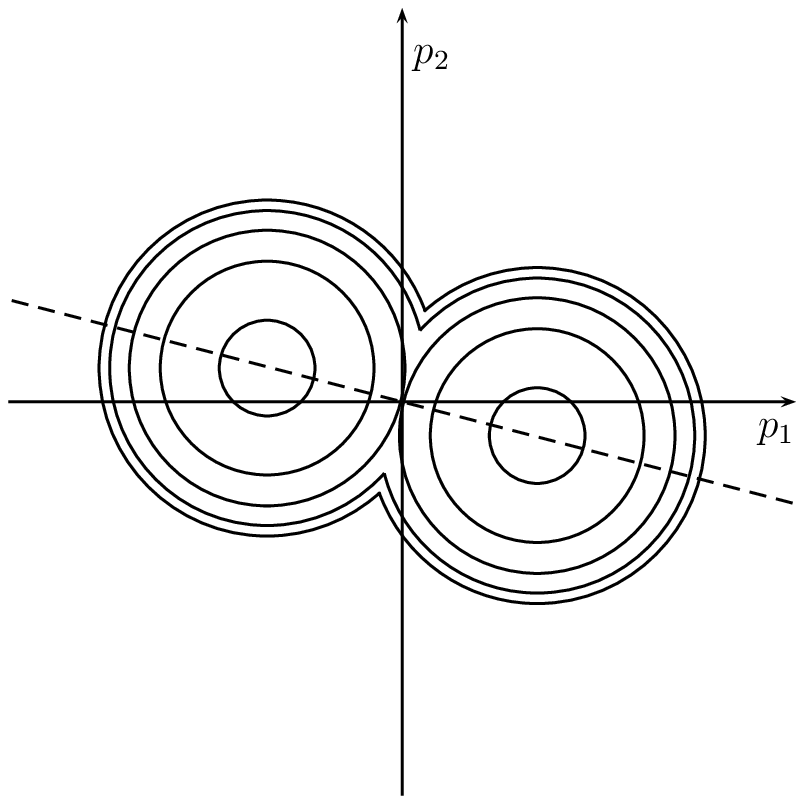,width=7cm} &
\epsfig{file=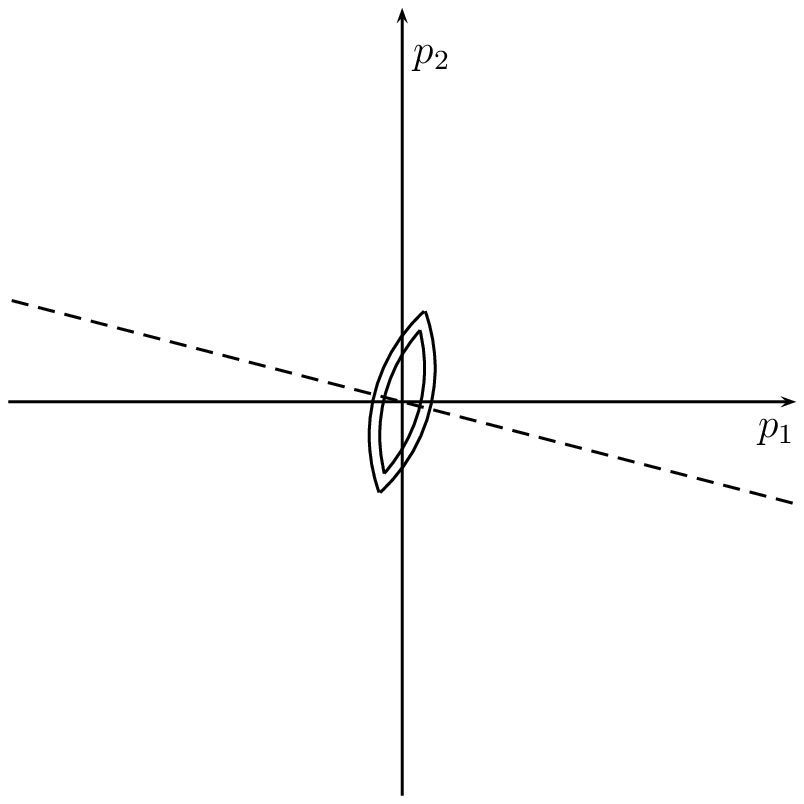,width=7cm} \end{tabular}
\end{center}
\caption{\it Lines of constant energy for electrons propagating in a spiral
configuration. The contours of the lower energy $E_-(\vec p)$ are shown on the 
left, and the contours of the higher energy $E_+(\vec p)$ are displayed on the
right.}
\end{figure}
For given $c_i \neq 0$ we now fill the lowest energy states with a small number
of electrons. The filled electron pockets are circles centered around
$\pm 2 K M' (c_1,- c_2)$ with a radius determined by the kinetic energy
\begin{equation}
T = \frac{1}{2 M'}\left[(p_1 \mp 2 K M' c_1)^2 + (p_2 \pm 2 K M' c_2)^2
\right]
\end{equation}
of an electron at the Fermi surface. The two occupied circular electron pockets
define a region $P$ in momentum space. The area of this region determines the
fermion density as
\begin{equation}
n = \frac{1}{(2 \pi)^2} \int_P d^2p = \frac{1}{\pi} M' T.
\end{equation}
The two circles do not overlap as long as $n < \frac{2}{\pi} M'^2 K^2 c_i c_i$.
The kinetic energy density of the filled region $P$ is given by
\begin{equation}
t = \frac{1}{(2 \pi)^2} \int_P d^2p \ 
\frac{1}{2 M'}\left[(p_1 \mp 2 K M' c_1)^2 + (p_2 \pm 2 K M' c_2)^2 \right] =
\frac{1}{2 \pi} M' T^2 = \frac{\pi n^2}{2 M'},
\end{equation}
and the total energy density of electrons then is
\begin{equation}
\epsilon_e = (M  + N c_i c_i - 2 K^2 M' c_i c_i) n + \frac{\pi n^2}{2 M'}.
\end{equation}
The resulting total energy density that includes the vacuum energy density
$\epsilon_0$ as well as the magnetic energy density $\epsilon_m$ is given by
\begin{equation}
\epsilon = \epsilon_0 + \epsilon_m + \epsilon_e = 
\epsilon_0 + 2 \rho_s c_i c_i + (M  + N c_i c_i - 2 K^2 M' c_i c_i) n + 
\frac{\pi n^2}{2 M'}.
\end{equation}
For $\rho_s > (K^2 M' - \frac{1}{2} N) n$ (which is always satisfied for
sufficiently small density $n$) the energy is minimized for $c_i = 0$ and the
value of the energy density at the minimum is given by
\begin{equation}
\label{etot}
\epsilon = \epsilon_0 + M n + \frac{\pi n^2}{2 M'}.
\end{equation}

However, one should not forget that, as the $c_i$ become smaller, the two 
occupied circles eventually touch each other once 
$\frac{2}{\pi} M'^2 K^2 c_i c_i = n$. Interestingly, in this moment the states 
with energy $E_+(\vec p)$ also become occupied. Indeed, as one can see in 
figure 6, the almond-shaped region
of occupied states with energy $E_+(\vec p)$ and the peanut-shaped region of
occupied states with energy $E_-(\vec p)$, combine to two complete overlapping
circles. This is also illustrated in figure 7, in which the energies 
$E_-(\vec p)$ and $E_+(\vec p)$ combine to form two overlapping parabolic 
dispersion relations.
\begin{figure}[t]
\begin{center}
\vspace{-0.4cm}
\epsfig{file=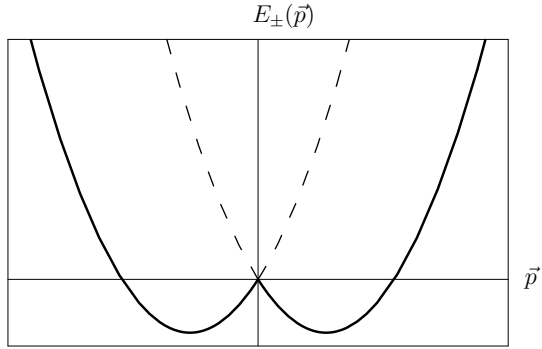,width=7cm}
\end{center}
\caption{\it The energies $E_-(\vec p)$ (solid curve) and $E_+(\vec p)$ (dashed
curve) along the line $\vec p \propto (c_1,- c_2)$ (the dashed lines in figure 
6) define two independent parabolic dispersion relations.}
\end{figure}
As a result, eq.(\ref{etot}) is still valid even when the occupied circles
overlap. Consequently, the energy minimum is indeed at $c_i = 0$ and thus a
homogeneous phase arises. This is in contrast to hole-doped cuprates for which
a spiral phase is energetically favored for intermediate values of the spin 
stiffness $\rho_s$ \cite{Bru06a}. The effective theory predicts that spiral 
phases are absent in electron-doped antiferromagnets.

\subsection{Inclusion of 4-Fermion Couplings}

Let us also calculate the effect of the 4-fermion contact interactions on the
energy density. We perform this calculation to first order of perturbation
theory, assuming that the 4-fermion interactions are weak. Depending on the 
underlying microscopic system such as the Hubbard model, the 4-fermion 
couplings may or may not be small.
We like to point out that, while the on-site Coulomb repulsion responsible for
antiferromagnetism is always large in the microscopic systems, the 4-fermion
couplings in the effective theory may still be small. If they are large, the 
result of the perturbative calculation should not be trusted. 

The perturbation of the Hamiltonian due to the leading 4-fermion contact term 
of eq.(\ref{Lagrange}) is given by
\begin{equation}
\label{DeltaH}
\Delta H = \frac{G_1}{2} \int d^2x \sum_{s = +,-}
\Psi^\dagger_s \Psi_s \Psi^\dagger_{-s} \Psi_{-s}.
\end{equation}
It should be noted that $\Psi^\dagger_s(x)$ and $\Psi_s(x)$ again are 
fermion creation and annihilation operators (and not Grassmann numbers).  
The terms proportional to $G_2, G_3,...,G_7$ are of higher order and will hence
not be taken into account. The fermion density is equally distributed among the
two spin orientations such that
\begin{equation}
\langle \Psi^\dagger_+ \Psi_+ \rangle = \langle \Psi^\dagger_- \Psi_- \rangle =
\frac{n}{2}.
\end{equation}
The brackets denote expectation values in the unperturbed state determined
before. Since the fermions are uncorrelated we have
\begin{equation}
\langle \Psi^\dagger_s \Psi_s \Psi^\dagger_{-s} \Psi_{-s} \rangle
= \langle \Psi^\dagger_s \Psi_s \rangle 
\langle \Psi^\dagger_{-s} \Psi_{-s}\rangle.
\end{equation}
Taking the 4-fermion contact terms into account in first order perturbation
theory, the total energy density of eq.(\ref{etot}) receives an additional 
contribution and now reads
\begin{equation}
\epsilon = \epsilon_0 + M n + \frac{\pi n^2}{2 M'} + \frac{G_1}{4} n^2.
\end{equation}

\section{Reduction of the Staggered Magnetization upon Doping}

The order parameter of the undoped antiferromagnet is the local staggered 
magnetization $\vec M_s(x) = {\cal M}_s\,\vec e(x)$ with ${\cal M}_s$ being the
length of the staggered magnetization vector. In a doped antiferromagnet the
staggered magnetization receives additional contributions from the electrons
such that
\begin{equation}
\vec M_s(x) = \Big[{\cal M}_s - 
m \sum_{s = +,-} \psi^\dagger_s(x) \psi_s(x) \Big] \vec e(x).
\end{equation}
The low-energy parameter $m$ determines the reduction of the staggered 
magnetization upon doping. Further contributions to $\vec M_s(x)$ which include
derivatives or contain more than two fermion fields are of higher order and
have thus been neglected. Using
\begin{equation}
\sum_{s = +,-} \langle \Psi^\dagger_s \Psi_s 
\rangle = n,
\end{equation}
we then obtain
\begin{equation}
{\cal M}_s(n) = {\cal M}_s - m n,
\end{equation}
i.e.\ at leading order the staggered magnetization decreases linearly with
increasing electron density. The higher-order terms that we have neglected will
give rise to sub-leading corrections of ${\cal O}(n^2)$.

\section{Conclusions}

In analogy to the hole-doped case \cite{Kae05,Bru06}, we have constructed a 
systematic effective field theory for lightly electron-doped antiferromagnets. 
Interestingly, the different locations of electron- and hole-pockets in the
Brillouin zone have important consequences for the dynamics. 

In the hole-doped case, the pockets are located at 
$(\pm \frac{\pi}{2 a},\pm \frac{\pi}{2 a})$ which gives rise to a flavor index 
that determines to which pocket a hole belongs. Due to spontaneous symmetry
breaking, holes and magnons are derivatively coupled. The leading magnon-hole
coupling contains a single spatial derivative and is responsible for a variety 
of interesting effects. First, it leads to a $1/r^2$ potential between a pair 
of holes which gives rise to an infinite number of two-hole bound states
\cite{Bru05,Bru06}. Remarkably, in the hole-doped case, in the 
$c \rightarrow \infty$ limit the symmetries give rise to an accidental
Galilean boost invariance. Hence, it is sufficient to consider the bound state
in its rest-frame. Second, in the hole-doped systems, the single-derivative 
magnon-hole coupling gives rise to a spiral phase for intermediate values of 
$\rho_s$.

In the electron-doped case discussed in this paper, the momentum space pockets
are located at $(\frac{\pi}{a},0)$ and $(0,\frac{\pi}{a})$. Due to 
antiferromagnetism these are actually two half-pockets which combine to a
single electron-pocket. Hence, in contrast to the hole-doped systems, electrons
do not carry an additional flavor index. As in the hole-case, electrons and 
magnons are derivatively coupled. However, due to the different implementation 
of the symmetries, the leading magnon-electron coupling now contains two 
spatial
derivatives. In other words, at low energies magnons are coupled to holes more
strongly than to electrons. As a consequence, the one-magnon exchange potential
between two electrons in their rest-frame decays as $1/r^4$ and is hence
weaker at large distances than in the hole-case. Still, magnon exchange is
capable of binding electrons. As another consequence of symmetry 
considerations, an accidental Galilean boost invariance is absent in the
electron-case. Indeed, the one-magnon exchange potential depends on the total
momentum $\vec P$ of the electron-pair, and it is hence not sufficient to 
consider the system in its rest-frame. The momentum-dependent contribution to
the potential is proportional to $P^2/r^2$, which gives rise to a non-trivial
structure of moving bound states. As another consequence of the weakness of the
magnon-electron coupling, in contrast to the hole-doped case, spiral phases are
energetically unfavorable for electron-doped systems. While this is not a new
result, we find it remarkable that it follows unambiguously from the very few 
basic assumptions of the systematic low-energy effective field theory, such as 
locality, symmetry, and unitarity.

We like to point out that the systematic effective field theory approach is
universally applicable to a large class of antiferromagnets. While it remains 
to be seen if the effective theory can also be applied to high-temperature 
superconductors, it makes unbiased, quantitative predictions for both lightly 
hole- and electron-doped cuprates and should be pursued further.

\section*{Acknowledgements}

C.\ P.\ H. would like to thank the members of the Institute for Theoretical 
Physics at Bern University for their hospitality. This work was supported in 
part by funds provided by the Schweizerischer Nationalfonds. The work of 
C.\ P.\ H.\ is supported by CONACYT grant No.\ 50744-F.

\begin{appendix}

\section{Magnon Exchange Potential in Coordinate \\ Space}

In this appendix we discuss the transformation of the one-magnon exchange
potential between two electrons from momentum space to coordinate space, which
is not entirely straightforward.

In momentum space the one-magnon exchange potential is given by
\begin{equation}
\langle \vec p_+ \ \!\!\!\! ' \ \vec p_- \ \!\!\!\! '|V
|\vec p_+ \vec p_-\rangle = V(\vec p,\vec p \ ') 
\delta(\vec p_+ + \vec p_- - \vec p_+ \ \!\!\!\! ' - \vec p_- \ \!\!\!\! '),
\end{equation}
with
\begin{equation}
V(\vec p,\vec p \ ') = \frac{K^2}{2 \rho_s} \frac{1}{q^2}
\left[q_1^2 - q_2^2 + 2(q_1 p_{-1} - q_2 p_{-2})\right]
\left[q_1^2 - q_2^2 - 2(q_1 p_{+1} - q_2 p_{+2})\right].
\end{equation}
Using
\begin{gather}
\vec P = \vec p_+ + \vec p_- = \vec p_+ \ \!\!\!\! ' + \vec p_- \ \!\!\!\! ',
\nonumber \\
\vec p = \frac{1}{2}(\vec p_+ - \vec p_-), \qquad
\vec p \ ' = \frac{1}{2}(\vec p_+ \ \!\!\!\! ' - \vec p_- \ \!\!\!\! '), \qquad
\vec q = \vec p + \vec  p \ ',
\end{gather}
it is easy to show that
\begin{equation}
V(\vec p,\vec p \ ') = V_0(\vec p,\vec p \ ') + V_{\vec P}(\vec p,\vec p \ '),
\end{equation}
with the rest-frame potential
\begin{equation}
V_0(\vec p,\vec p \ ') = \frac{K^2}{2 \rho_s} 
\frac{p_1^2 - {p_1'}^2 - p_2^2 + {p_2'}^2}{(p_1 + p_1')^2 + (p_2 + p_2')^2},
\end{equation}
and the momentum-dependent contribution
\begin{equation}
V_{\vec P}(\vec p,\vec p \ ') = - \frac{K^2}{2 \rho_s} 
\frac{\left[P_1 (p_1 + p_1') - P_2 (p_2 + p_2')\right]}
{(p_1 + p_1')^2 + (p_2 + p_2')^2}.
\end{equation}
The potential in coordinate space is the Fourier transform of the potential in
momentum space
\begin{equation}
V(\vec x,\vec x \ ') = \frac{1}{(2 \pi)^4} \int d^2p \ d^2p' \ 
V(\vec p,\vec p \ ') 
\exp(i \vec p \cdot \vec x - i \vec p \ ' \cdot \vec x \ ').
\end{equation}
Introducing
\begin{equation}
\vec k = \frac{1}{2}(\vec p - \vec p \ '), \qquad
\vec r = \frac{1}{2}(\vec x - \vec x \ '), \qquad
\vec y = \vec x + \vec x \ ',
\end{equation}
one obtains
\begin{equation}
\vec p \cdot \vec x - \vec p \ ' \cdot \vec x \ ' = 
\vec q \cdot \vec r + \vec k \cdot \vec y,
\end{equation}
such that the momentum-dependent contribution takes the form
\begin{equation}
V_{\vec P}(\vec x,\vec x \ ') = - \frac{K^2}{2 \rho_s} \frac{1}{(2 \pi)^2}
\int d^2q \ \frac{\left(P_1 q_1 - P_2 q_2\right)^2}{q_1^2 + q_2^2}\, 
\exp(i \vec q \cdot \vec r)\, \delta(\vec y).
\end{equation}
The $\delta$-function arises from the $k$-integration and implies 
$\vec x \ ' = - \vec x$ as well as $\vec r = \vec x$, which just means that the
potential is local in coordinate space. Using
\begin{equation}
\frac{1}{(2 \pi)^2} \int d^2q \ \frac{q_1^2 - q_2^2}{q_1^2 + q_2^2} =
- \frac{1}{\pi} \frac{\cos(2 \varphi)}{r^2}, \qquad
\frac{1}{(2 \pi)^2} \int d^2q \ \frac{2 q_1 q_2}{q_1^2 + q_2^2} =
- \frac{1}{\pi} \frac{\sin(2 \varphi)}{r^2},
\end{equation}
with $\vec r = r (\cos\varphi,\sin\varphi)$ the $q$-integration results in
\begin{align}
V_{\vec P}(\vec x,\vec x \ ')&=\frac{K^2}{2 \pi \rho_s} 
\left[\frac{1}{2}(P_1^2 - P_2^2) \frac{\cos(2 \varphi)}{r^2} - P_1 P_2
\frac{\sin(2 \varphi)}{r^2}\right] \delta(\vec y) \nonumber \\[.5ex]
&=\frac{K^2 P^2}{4 \pi \rho_s}\, \frac{\cos\big(2 (\varphi + \chi)\big)}{r^2}\,
\delta(\vec y).  
\end{align}
In the last step we have introduced $\vec P = P (\cos\chi,\sin\chi)$.

Similarly, the rest-frame potential takes the form
\begin{equation}
\label{V_0}
V_0(\vec x,\vec x \ ') = \frac{K^2}{2 \rho_s} \frac{1}{(2 \pi)^4}
\int d^2q \ d^2k \ \frac{\left(2 q_1 k_1 - 2 q_2 k_2\right)^2}{q_1^2 + q_2^2} 
\exp(i \vec q \cdot \vec r) \exp(i \vec k \cdot \vec y).
\end{equation}
The $k$-integration results in the second derivative of a $\delta$-function 
which again implies $\vec x \ ' = - \vec x$ as well as $\vec r = \vec x$. 
Hence, also the rest-frame potential is local and one can write
\begin{equation}
V_0(\vec x,\vec x \ ') = V_{ij}(\vec r) \p_{y_i} \p_{y_j} \delta(\vec y),
\end{equation}
with $V_{ij}(\vec r)$ implicitly defined through eq.(\ref{V_0}). In order to
figure out how $V_0(\vec x,\vec x \ ')$ acts on a wave function we calculate
\begin{align}
\langle \Phi|V_0|\Psi \rangle&=\int d^2x \ d^2x' \ \langle \Phi|\vec x \rangle
V_0(\vec x,\vec x \ ') \langle \vec x \ '|\Psi\rangle \nonumber \\
&=\int d^2x \ d^2x' \ \langle \Phi|\vec x \rangle
V_{ij}(\vec r) \p_{y_i} \p_{y_j} \delta(\vec y) \langle \vec x \ '|\Psi\rangle 
\nonumber \\
&=\int d^2r \ d^2y \ \langle \Phi|\frac{\vec y}{2} + \vec r \rangle
V_{ij}(\vec r) \p_{y_i} \p_{y_j} \delta(\vec y) 
\langle \frac{\vec y}{2} - \vec r|\Psi\rangle 
\nonumber \\
&=\frac{1}{4} \int d^2r \ V_{ij}(\vec r) \p_{r_i} \p_{r_j} 
(\langle \Phi|\vec r \rangle \langle - \vec r|\Psi\rangle) \nonumber \\
&=\frac{1}{4} \int d^2r \ \langle \Phi|\vec r \rangle 
\left[\p_{r_i} \p_{r_j} V_{ij}(\vec r)\right] \langle - \vec r|\Psi\rangle.
\end{align}
It is now straightforward to convince oneself that
\begin{equation}
\frac{1}{4} \p_{r_i} \p_{r_j} V_{ij}(\vec r) = 
\frac{6 K^2}{\pi \rho_s} \frac{r_1^4 - 6 r_1^2 r_2^2 + r_2^4}{r^8} =
\frac{6 K^2}{\pi \rho_s} \frac{\cos(4 \varphi)}{r^4}.
\end{equation}

Altogether, in coordinate space the resulting potential is hence given by
\begin{equation}
\langle \vec r_+ \ \!\!\!\! ' \vec r_- \ \!\!\!\! '|V
|\vec r_+ \vec r_-\rangle = \frac{K^2}{2 \pi \rho_s} 
\left[12 \frac{\cos(4 \varphi)}{r^4} + 
\frac{P^2}{2} \frac{\cos\big(2 (\varphi + \chi)\big)}{r^2}
\right] \delta(\vec r_+ - \vec r_- \ \!\!\!\! ') \
\delta(\vec r_- - \vec r_+ \ \!\!\!\! ').
\end{equation}

\end{appendix}

\end{document}